# Spatial-Temporal Imaging of Anisotropic Photocarrier Dynamics in Black Phosphorus

Bolin Liao[1,2,⊥], Huan Zhao[3,⊥], Ebrahim Najafi[1], Xiaodong Yan[3], He Tian[3], Jesse Tice[4], Austin J. Minnich[2,5]\*, Han Wang[3]\* and Ahmed H. Zewail[1,2,§]

[1]Division of Chemistry and Chemical Engineering, California Institute of Technology, Pasadena, CA 91125, USA
[2]Kavli Nanoscience Institute, California Institute of Technology, Pasadena, CA 91125
[3]Ming Hsieh Department of Electrical Engineering, University of Southern California, Los Angeles, CA 90089, USA
[4]NG Next, Northrop Grumman, 1 Space Park, Redondo Beach, CA 90278, USA
[5]Division of Engineering and Applied Science, California Institute of Technology, Pasadena, CA 91125, USA

**As an emerging single elemental layered material with a low symmetry in-plane crystal lattice[1-6], black phosphorus (BP) has attracted significant research interest owing to its unique electronic and optoelectronic properties, including its widely tunable bandgap[7-9], polarization dependent photoresponse[3,10] and highly anisotropic in-plane charge transport[3]. Despite extensive study of the steady-state charge transport in BP[2-5,11,12], there has not been direct characterization and visualization of the hot carriers dynamics in BP immediately after photoexcitation, which is crucial to understanding the performance of BP-based optoelectronic devices. Here we use the newly developed scanning ultrafast electron microscopy (SUEM)[13,14] to directly visualize the motion of photo-excited hot carriers on the surface of BP in both space and time. We observe highly anisotropic in-plane diffusion of hot holes, with a 15-times higher diffusivity along the armchair (x-) direction than that along the zigzag (y-) direction. Our results provide direct evidence of anisotropic hot carrier transport in BP and demonstrate the capability of SUEM to resolve ultrafast hot carrier dynamics in layered two-dimensional materials.**

[⊥]These authors contributed equally.
\*To whom correspondence should be addressed: han.wang.4@usc.edu (H.W.), aminnich@caltech.edu  (A.J.M.)
[§] Deceased

With a widely tunable band gap from 0.3 eV in its bulk form to above 1.3 eV as a single layer, black phosphorus (BP)[1-6] has recently emerged as a promising candidate material for infrared optoelectronic applications,[10,15,16] due to its highly tunable and customizable electronic[7-9,17] and optical properties. In addition, the relatively high charge mobility of BP adds to its attractiveness for applications that require efficient charge transport.[18] One of the most intriguing features of BP, however, is its strong in-plane anisotropy,[3,4] originating from the low-symmetry puckered orthorhombic lattice structure. The near-equilibrium charge mobility along the armchair direction is known to be higher than that along the zigzag direction, revealed by both the field-effect and Hall measurements.[2-4] Due to the opposite trend in its thermal conductivity,[19-21] the strong anisotropy of BP is also believed to be promising for thermoelectric applications.[22] Moreover, the in-plane anisotropy leads to strongly polarization-dependent optical properties, rendering BP a suitable material for polarization-sensitive detectors.[10,16]

Besides near-equilibrium transport, another important aspect of charge transport is the dynamics of photo-excited hot carriers relevant to the photo-detection and photovoltaic applications. Due to the initial high temperature of the photo-excited charge carriers, the hot-carrier dynamics can be drastically different from near-equilibrium transport. Therefore, a thorough understanding in the motion of hot electrons and holes in BP immediately after photo-excitation is essential for designing and improving BP-based optoelectronic devices. Previous studies[23-26] exclusively utilized ultrafast optical pump-probe spectroscopy to investigate the transient change of absorption or transmission of BP induced by the pumping laser pulse, from which the dynamics of hot carriers could be indirectly inferred. Restricted by the optical diffraction limit, however, optical pump-probe spectroscopy lacks the spatial resolution to directly map out the diffusion process of photo-excited charge carriers.

Scanning ultrafast electron microscopy (SUEM)[13,14] is a newly developed technique that can directly image the dynamics of photo-excited carriers in both space and time, with sub-picosecond temporal resolution and nanometer spatial resolution. Details of the setup can be found elsewhere[13,14,27-29] and are briefly summarized here (also illustrated in Figure 1a). Compared to optical pump-probe spectroscopy, SUEM is a photon-pump-electron-probe technique, with sub-picosecond electron pulses generated by illuminating a photocathode (ZrO-coated tungsten tip) with an ultrafast ultraviolet (UV) laser beam (wavelength 257 nm, pulse duration 300 fs, repetition rate 5 MHz, fluence 300 $\mu J/cm^2$). A typical probing electron pulse consists of tens to hundreds of electrons, estimated by measuring the beam current through a Faraday cup. The probing electron pulses arrive at the sample after the optical pump pulses (wavelength 515 nm, fluence 80 $\mu J/cm^2$) by a given time controlled by a mechanical delay stage (-700 ps to 3.6 ns, with 1 ps resolution). The probing electron pulses induce the emission of secondary electrons from the sample, which are subsequently collected by an Everhart-Thornley detector. To form an image, the probing electron pulses are scanned across the sample surface and the secondary electrons emitted from each location are counted. Since the yield of secondary electrons depends on the local average electron energy, more/less secondary electrons are emitted from regions of the sample surface where there is a net accumulation of electrons/holes.[27] Typically a reference SEM image is taken long before the pump optical pulse arrives and is then subtracted from images taken at other delay times to remove the background. In the resulting "contrast images", bright/dark contrasts are observed at places with net accumulation of electrons/holes due to higher/lower yield of secondary electrons. In this fashion, the dynamics of electrons and holes after excitation by the optical pump pulse can be monitored in real space and time.[14] An alternative way of visualizing hot carrier dynamics in space and time was recently demonstrated using time-resolved photoemission electron microscopy (TR-PEEM).[30]

In this letter, we demonstrate the direct imaging of hot-carrier dynamics in BP with SUEM. Due to the presence of a surface potential, we observe the motion of hot holes on the surface of BP. With SUEM, we see striking visualization of anisotropic diffusion of hot holes after photo-excitation, from which quantitative transport parameters can be extracted. Our results indicate a 15-times higher diffusivity of hot-holes moving along the armchair direction than that along the zigzag direction, which is a combined effect of anisotropic effective mass and direction-independent electron-phonon scattering.[31]

Figure 1b displays a static SEM image of a typical BP flake measured in this work. BP flakes of 80 nm thickness were mechanically exfoliated from a bulk crystal, and subsequently transferred to an ITO-coated glass substrate to avoid charging of the sample. The sample is exclusively handled in an argon glovebox before being immediately loaded into the SUEM vacuum chamber. The arrows in Figure 1b denote the armchair (x-) and zigzag (y-) directions of the BP crystal, determined optically by Raman spectroscopy.[32] Polarization-resolved Raman spectroscopy was performed after the SUEM measurements using a 532 nm Nd:YAG laser in the LabRAM ARAMIS system. A 100× microscope objective was used and the power incident on the BP sample was kept below 500 µW to avoid sample damage. Polarization resolved Raman spectroscopy was conducted with a 532 nm laser with different sample rotating angles. The dependence of the Raman peaks intensity on the rotation angle of the sample basal plane is shown in Figure 1c. The crystal orientation is determined specifically from the intensity of the $A_g^1$ peak, which reaches the minimal intensity when the laser polarization is along armchair (x-) direction.[32]

SUEM contrast images of the BP flake shown in Figure 1b are presented in Figure 2. A low-pass Gaussian filter is used to suppress the noise of the images for presentation, while raw images are used for quantitative analysis shown later. Images displayed in the first row were taken when the flake is oriented as shown in Figure 1b, whereas images in the second row were taken when the flake is rotated by 90 degree. Firstly, only dark contrast is observed in the region of the sample excited by the pump laser. The initial shape of the excited spot is elliptical as expected because the pump laser is incident on the sample at an angle. Since secondary electrons are typically emitted only from the top few nanometers of the sample, the observation of only dark contrast indicates that the electrons and holes are separated vertically after photo-excitation, and the holes are accumulated near the sample surface while the electrons are drawn away from the surface. This separation is most likely due to the existence of a surface potential on the BP sample (we observed the same behavior in heavily doped n-type silicon with SUEM also caused by a surface potential), which arises due to the formation of an atomically thin phosphorus oxide layer[33] on the BP surface when the samples are briefly exposed to air (<30 seconds) during their loading into the vacuum chamber of the SUEM. The vertical transport process associated with the surface potential is also reflected in the observation that the intensity of the dark contrast reaches a maximum around 40 ps after the optical pump pulse arrives. At 40 ps delay time, the profile of the spatial distribution of holes follows approximately the shape of the pump beam. As the time progresses, it is clearly observed that the holes preferentially diffuse along the armchair (x-) direction, denoted by the orange arrow, regardless of the relative orientation of the BP flake and the optical pump beam. In this measurement the polarization of the optical pump beam is not specifically chosen. Although it is known that the absorption of BP is strongly dependent on the light polarization,[10] we verified experimentally that the light polarization does not affect the dynamics of the hot holes after photo-excitation, as shown in Supplementary Figure 2.

In addition to the direct and intuitive visualization of the highly anisotropic transport of photo-excited hot holes in BP provided by the SUEM contrast images in Figure 2, numerical values of transport parameters can be extracted through quantitative analysis of the contrast images. A convenient parameter to describe the spatial distribution of particles is the variance $\sigma$, which in this case is angle-dependent, defined as

$$\sigma(\theta,t) = \frac{\iint \rho(\mathbf{r},t)(\mathbf{r}\cdot\hat{\theta})^2 d^2\mathbf{r}}{\iint \rho(\mathbf{r},t) d^2\mathbf{r}} - \left(\frac{\iint \rho(\mathbf{r},t)(\mathbf{r}\cdot\hat{\theta}) d^2\mathbf{r}}{\iint \rho(\mathbf{r},t) d^2\mathbf{r}}\right)^2, \quad (1)$$

where $\rho(\mathbf{r},t)$ is the local carrier concentration, and $\hat{\theta}$ is a unit vector pointing to a certain angle $\theta$. Here we approximate $\rho(\mathbf{r},t)$ with the measured local intensity $I(\mathbf{r},t)$ of the SUEM contrast images, assuming a linear relation between them. This assumption should hold here since the optical excitation is weak and the measurement works in the linear response regime. Figure 3a shows the calculated $\sigma(\theta,t)$ from the images at different delay times in the first row of Figure 2, normalized by $\sigma(\theta,t)$ at 40 ps. It is clear that the variance of the spatial distribution increases significantly along the armchair direction, while its change along the zigzag direction is hardly discernible.

In the case when the diffusivity is a constant, the variance of the spatial distribution should be a linear function of time. In the current experiment, however, the diffusivity changes with time due to a decreasing temperature of the hot carriers. As a first-order approximation, we assume the temperature of the hot carriers decays exponentially $T(t) = T_0 \exp(-t/\tau_T)$, with a time constant $\tau_T$ controlled mainly by inelastic electron-phonon scatterings.[34] We further apply the Einstein relation $D = (k_B T \mu)/q$, where $D$ is the diffusivity and $\mu$ is the mobility, and argue that $D(t) = D_0 \exp(-t/\tau_T)$, where $D_0$ is the diffusivity right after photo-excitation, because immediately after photo-excitation, the lattice is still cold so that the mobility limited by electron-phonon interaction is not largely affected. In this case the time dependence of the variance is:

$$\sigma(\theta,t) = 2D_0(\theta)\tau_T \left[1 - \exp\left(-\frac{t}{\tau_T}\right)\right] + \sigma_0(\theta). \qquad (2)$$

Fits of Eq. (2) to experimentally measured variances along the armchair and zigzag directions are plotted in Figure 3b. From the fittings, the parameters can be extracted as $\tau_T \approx 150$ ps, $D_{0,\text{armchair}} \approx 1.3 \times 10^4$ cm$^2$/s, $D_{0,\text{zigzag}} \approx 870$ cm$^2$/s, $\sigma_{0,\text{armchair}} \approx 216.7$ $\mu$m$^2$ and $\sigma_{0,\text{zigzag}} \approx 405.8$ $\mu$m$^2$. The ratio between the diffusivities along the two directions is approximately 15. This ratio is much higher than that measured by steady-state transport experiments[4] and calculated by first-principles simulations assuming near-equilibrium transport,[31] but close to the value inferred from an optical pump-probe measurement.[26] This observation demonstrates the significant difference between hot-carrier dynamics and near-equilibrium dynamics, and the fact that the measured hot-carrier diffusivity ratio is on the same order of the effective mass ratio[35] indicates that the transport of photo-excited hot carriers is likely more affected by the effective mass of the carriers than their scattering properties. Furthermore, the timescale of carrier recombination can be inferred from the time-dependence of the average intensity of the dark contrast, as shown in Fig. 3c. Data collected from three different BP samples are compiled here, each covering a different range of delay time, and the exponential fit gives a recombination lifetime $\tau_R$ ~550 ps. Within this recombination time, the average diffusion length of holes can be estimated, using Eq. (2), to be 9.7 μm along the armchair direction and 0.6 μm along the zigzag direction.

To further examine the validity of the extracted model parameters, here we simulate the SUEM contrast images by numerically solving the two-dimensional diffusion equation with time-dependent anisotropic diffusivities and a recombination term:

$$\frac{\partial \rho(\mathbf{r},t)}{\partial t} = D_x(t)\frac{\partial^2 \rho(\mathbf{r},t)}{\partial x^2} + D_y(t)\frac{\partial^2 \rho(\mathbf{r},t)}{\partial y^2} - \frac{\rho(\mathbf{r},t)}{\tau_R}, \tag{3}$$

where $x$ and $y$ directions are the armchair and zigzag directions, respectively, and $D_x$ and $D_y$ are time-dependent diffusivities along the two directions, with values as discussed in the previous sections. The simulated images are shown in Figure 4. The intensity of these simulated images represents the spatial concentration of holes normalized to the maximum value (at the center) of the initial distribution. The time-dependence of the distribution profile of hot holes in the simulation is in good agreement with the experimental images shown in the first row of Fig. 2, justifying the transport parameters extracted from our analysis of the experimental SUEM images.

In summary, we use SUEM to directly visualize the dynamics of photo-excited hot holes on the surface of black phosphorus. The highly anisotropic in-plane charge transport of black phosphorus is confirmed in our experiment, and we further find that the ratio between the diffusivities of hot holes along the armchair direction and the zigzag direction is much larger than that measured at near-equilibrium conditions, illustrating the drastic difference between hot carrier dynamics and near-equilibrium carrier dynamics. This study demonstrates the capability of SUEM in deepening our understanding of hot carrier dynamics in low-symmetry layered materials.

**Author contributions**

B.L., H.Z. and H.W. conceived the project. H.Z. prepared the BP samples. H.Z. and J.T. performed the Raman characterization. B.L. and E.N. carried out the SUEM measurements. B.L., H.Z., X.Y., H.T., A.J.M. and H.W. analyzed the data. B.L., H.Z., A.J.M. and H.W. wrote the manuscript. A.H.Z. led the development of the SUEM technique and supervised the research effort during the initial stage of the project.

**Competing financial interests**

Authors declare no competing financial interests.

**Acknowledgement**

This work is partially supported by the National Science Foundation (DMR-0964886) and the Air Force Office of Scientific Research (FA9550-11-1-0055) in the Gordon and Betty Moore Center for Physical Biology at the California Institute of Technology. The work is also supported by the Army Research Office (W911NF-16-1-0435), the Air Force Office of Scientific Research FATE MURI program (FA9550-15-1-0514) and the Northrop Grumman Institute of Optical Nanomaterials and Nanophotonics (NG-ION$^2$) at University of Southern California. B. L. is grateful for the financial support from the KNI Prize Postdoctoral Fellowship in Nanoscience at the Kavli Nanoscience Institute of California Institute of Technology.

**Reference**

1. Ling, X., Wang, H., Huang, S., Xia, F. & Dresselhaus, M. S. The renaissance of black phosphorus. *Proceedings of the National Academy of Sciences* **112**, 4523-4530 (2015).
2. Li, L. *et al.* Black phosphorus field-effect transistors. *Nature nanotechnology* **9**, 372-377 (2014).
3. Xia, F., Wang, H. & Jia, Y. Rediscovering black phosphorus as an anisotropic layered material for optoelectronics and electronics. *Nature communications* **5** (2014).
4. Liu, H. *et al.* Phosphorene: An Unexplored 2D Semiconductor with a High Hole Mobility. *ACS Nano* **8**, 4033-4041, doi:10.1021/nn501226z (2014).
5. Koenig, S. P., Doganov, R. A., Schmidt, H., Castro Neto, A. & Özyilmaz, B. Electric field effect in ultrathin black phosphorus. *Applied Physics Letters* **104**, 103106 (2014).
6. Castellanos-Gomez, A. *et al.* Isolation and characterization of few-layer black phosphorus. *2D Materials* **1**, 025001 (2014).
7. Rodin, A., Carvalho, A. & Neto, A. C. Strain-induced gap modification in black phosphorus. *Physical review letters* **112**, 176801 (2014).
8. Kim, J. *et al.* Observation of tunable band gap and anisotropic Dirac semimetal state in black phosphorus. *Science* **349**, 723-726 (2015).
9. Deng, B. *et al.* Efficient Electrical Control of Thin-Film Black Phosphorus Bandgap. *arXiv preprint arXiv:1612.04475* (2016).
10. Yuan, H. *et al.* Polarization-sensitive broadband photodetector using a black phosphorus vertical p–n junction. *Nature nanotechnology* **10**, 707-713 (2015).
11. Chen, X. *et al.* High-quality sandwiched black phosphorus heterostructure and its quantum oscillations. *Nature communications* **6** (2015).
12. Li, L. *et al.* Quantum Hall effect in black phosphorus two-dimensional electron system. *Nature nanotechnology* (2016).
13. Yang, D.-S., Mohammed, O. F. & Zewail, A. H. Scanning ultrafast electron microscopy. *Proceedings of the National Academy of Sciences* **107**, 14993-14998 (2010).
14. Najafi, E., Scarborough, T. D., Tang, J. & Zewail, A. Four-dimensional imaging of carrier interface dynamics in pn junctions. *Science* **347**, 164-167 (2015).
15. Youngblood, N., Chen, C., Koester, S. J. & Li, M. Waveguide-integrated black phosphorus photodetector with high responsivity and low dark current. *Nature Photonics* (2015).
16. Guo, Q. *et al.* Black phosphorus mid-infrared photodetectors with high gain. *Nano letters* **16**, 4648-4655 (2016).
17. Li, L. *et al.* Direct observation of the layer-dependent electronic structure in phosphorene. *Nature Nanotechnology* **12**, 21-25 (2017).
18. Wang, H. *et al.* Black phosphorus radio-frequency transistors. *Nano letters* **14**, 6424-6429 (2014).
19. Lee, S. *et al.* Anisotropic in-plane thermal conductivity of black phosphorus nanoribbons at temperatures higher than 100 K. *Nature communications* **6** (2015).
20. Smith, B. *et al.* Temperature and Thickness Dependences of the Anisotropic In‐Plane Thermal Conductivity of Black Phosphorus. *Advanced Materials* (2016).

21. Zhu, J. *et al.* Revealing the Origins of 3D Anisotropic Thermal Conductivities of Black Phosphorus. *Advanced Electronic Materials* **2** (2016).
22. Fei, R. *et al.* Enhanced thermoelectric efficiency via orthogonal electrical and thermal conductances in phosphorene. *Nano letters* **14**, 6393-6399 (2014).
23. Ge, S. *et al.* Dynamical evolution of anisotropic response in black phosphorus under ultrafast photoexcitation. *Nano letters* **15**, 4650-4656 (2015).
24. Wang, K. *et al.* Ultrafast nonlinear excitation dynamics of black phosphorus nanosheets from visible to mid-infrared. *ACS nano* **10**, 6923-6932 (2016).
25. Suess, R. J., Jadidi, M. M., Murphy, T. E. & Mittendorff, M. Carrier dynamics and transient photobleaching in thin layers of black phosphorus. *Applied Physics Letters* **107**, 081103 (2015).
26. He, J. *et al.* Exceptional and anisotropic transport properties of photocarriers in black phosphorus. *ACS nano* **9**, 6436-6442 (2015).
27. Cho, J., Hwang, T. Y. & Zewail, A. H. Visualization of carrier dynamics in p (n)-type GaAs by scanning ultrafast electron microscopy. *Proceedings of the National Academy of Sciences* **111**, 2094-2099 (2014).
28. Liao, B., Najafi, E., Li, H., Minnich, A. J. & Zewail, A. H. Dynamics of Photo-excited Hot Carriers in Hydrogenated Amorphous Silicon Imaged by 4D Electron Microscopy. *arXiv preprint arXiv:1610.03030* (2016).
29. Sun, J., Adhikari, A., Shaheen, B. S., Yang, H. & Mohammed, O. F. Mapping Carrier Dynamics on Material Surfaces in Space and Time using Scanning Ultrafast Electron Microscopy. *The journal of physical chemistry letters* **7**, 985-994 (2016).
30. Man, M. K. *et al.* Imaging the motion of electrons across semiconductor heterojunctions. *Nature Nanotechnology* **12**, 36-40 (2017).
31. Liao, B., Zhou, J., Qiu, B., Dresselhaus, M. S. & Chen, G. Ab initio study of electron-phonon interaction in phosphorene. *Physical Review B* **91**, 235419 (2015).
32. Phaneuf-L'Heureux, A.-L. *et al.* Polarization-resolved Raman study of bulk-like and Davydov-induced vibrational modes of exfoliated Black Phosphorus. *Nano Letters* **16**, 7761-7767 (2016).
33. Edmonds, M. *et al.* Creating a stable oxide at the surface of black phosphorus. *ACS applied materials & interfaces* **7**, 14557-14562 (2015).
34. Chen, G. *Nanoscale energy transport and conversion: a parallel treatment of electrons, molecules, phonons, and photons*. (Oxford University Press, 2005).
35. Qiao, J., Kong, X., Hu, Z.-X., Yang, F. & Ji, W. High-mobility transport anisotropy and linear dichroism in few-layer black phosphorus. *Nature communications* **5** (2014).

**Figure Captions**

**Figure 1. SUEM setup and BP sample.** (a) Schematic of the experimental setup. SE: secondary electron. (b) SEM image of a typical BP flake used for the SUEM measurements. The orange and yellow arrows denote the armchair (x-) and zigzag (y-) directions of the BP crystal, respectively, as determined by optical Raman measurement. Scale bar: 100 μm. (c) Raman characterization of the BP flake in (b). Left: the Raman spectra measured with the incident laser at different polarization angles. x denotes the armchair direction and y denotes the zigzag direction. Right: the intensity of the $A_g^1$ peak with the incident laser at different polarization angles.

**Figure 2. SUEM imaging of hole diffusion on the BP surface.** First row: the sample orientation is the same as that shown in Fig. 1(b). Second row: the sample is rotated by 90 degrees. The arrows denote the armchair direction. Scale bar: 60 μm. The orange and yellow arrows shown in the two left images denote the armchair (x-) and zigzag (y-) directions of the BP crystal for the original sample orientation (first row) and after the sample is rotated by 90 degrees (second row), respectively. A low-pass Gaussian filter is used to suppress the noise in the images for presentation. The dashed red ellipses are to guide the eye.

**Figure 3. Analysis of the SUEM images and hot carrier transport.** (a) The variance of hole-distribution along different directions, normalized by that of the initial distribution. (b) The variation of hole distribution along the armchair and zigzag directions versus time delay. Dashed lines represent fit with an exponentially decaying diffusivity due to the cooling process of the holes. (c) The average intensity of the hole-distribution versus time delay, indicating the time scale of the carrier recombination process. Data collected from three different BP samples are compiled here, each covering a different range of delay time. Dashed line is an exponential fit with a recombination time of 550 ps.

**Figure 4. Simulated anisotropic carrier diffusion in BP.** These images are simulated by numerically integrating Eq. (3) in the main text up to the corresponding delay time. Scale bar: 60 μm. The initial distribution at 40 ps is assumed to be a Gaussian of radius set by the incident pump laser beam. The intensity of these simulated images represents the spatial concentration of holes normalized to the maximum value (at the center) of the initial distribution.

# Spatial-Temporal Imaging of Anisotropic Photocarrier Dynamics in Black Phosphorus

# Figure 1

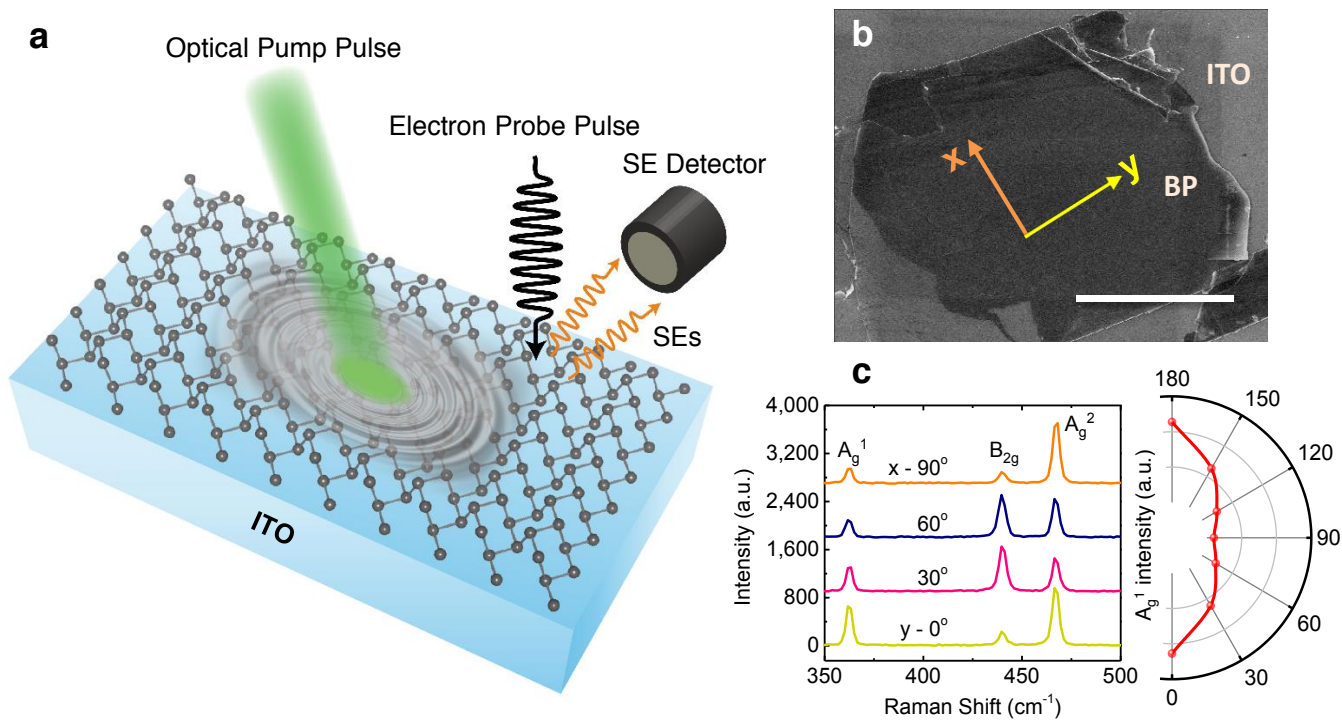

**Figure 2**

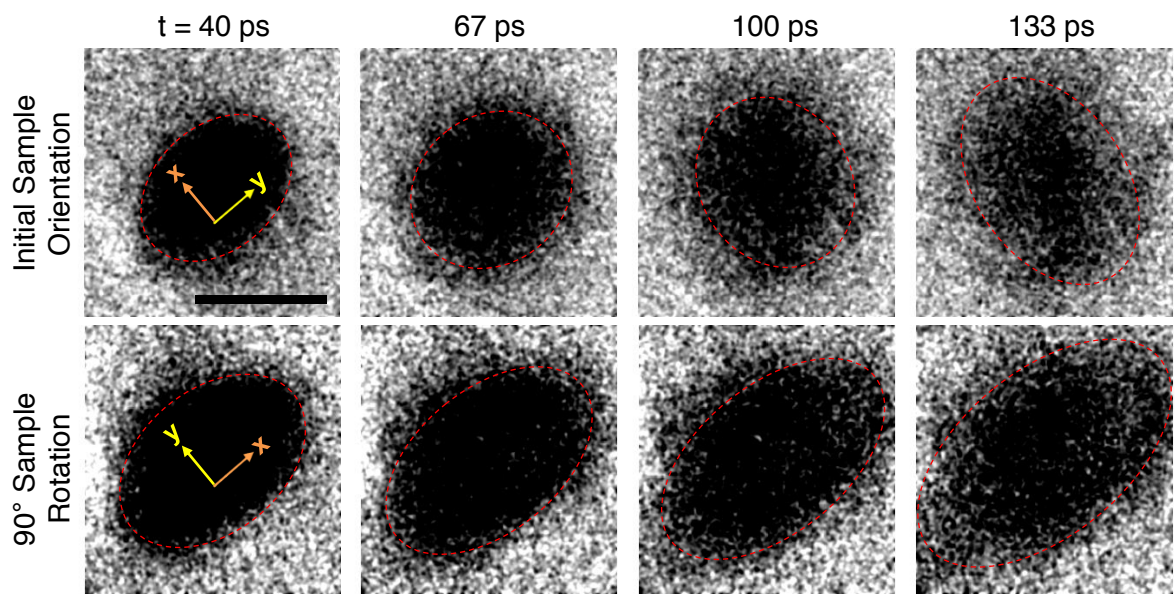

**Figure 3**

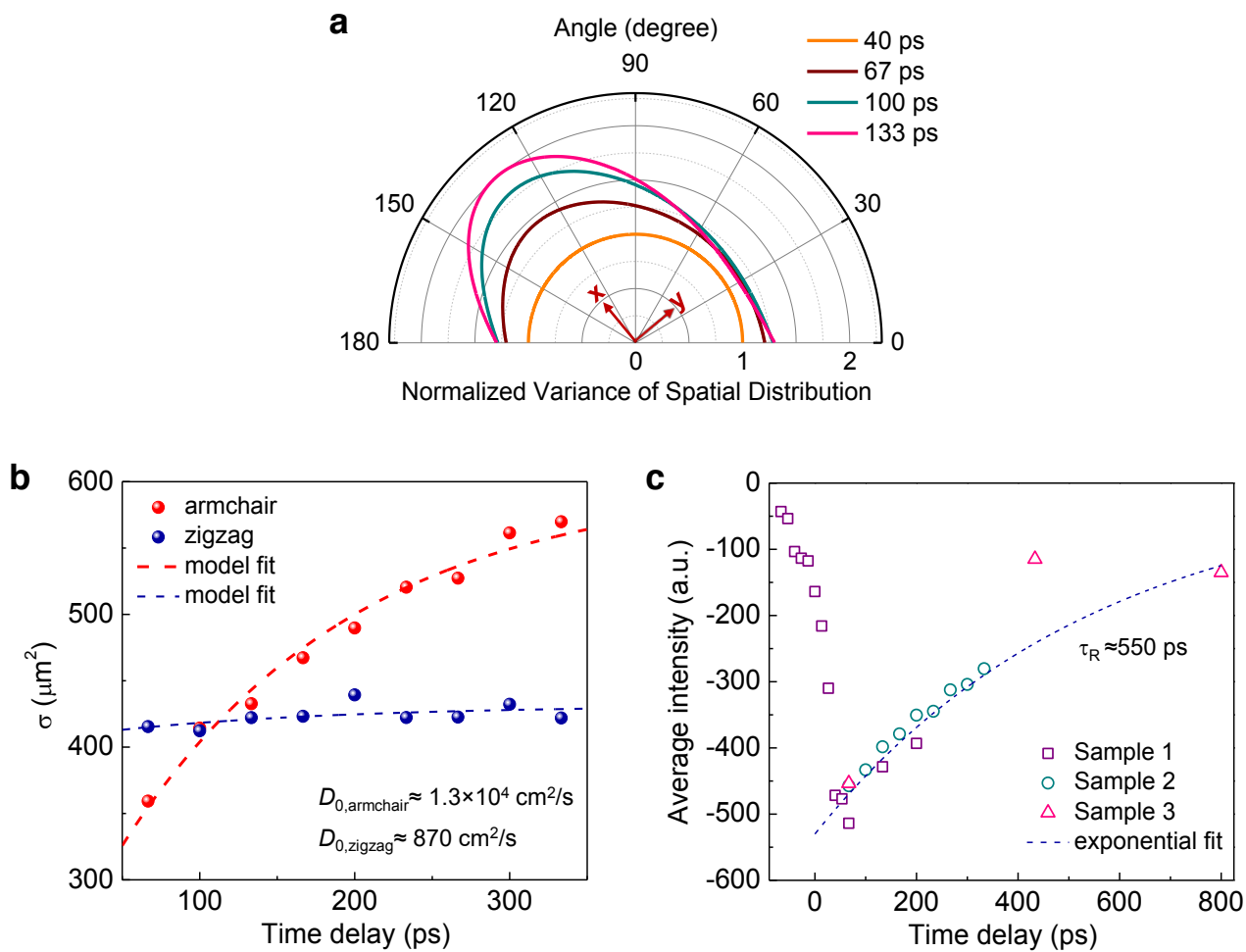

**Figure 4**

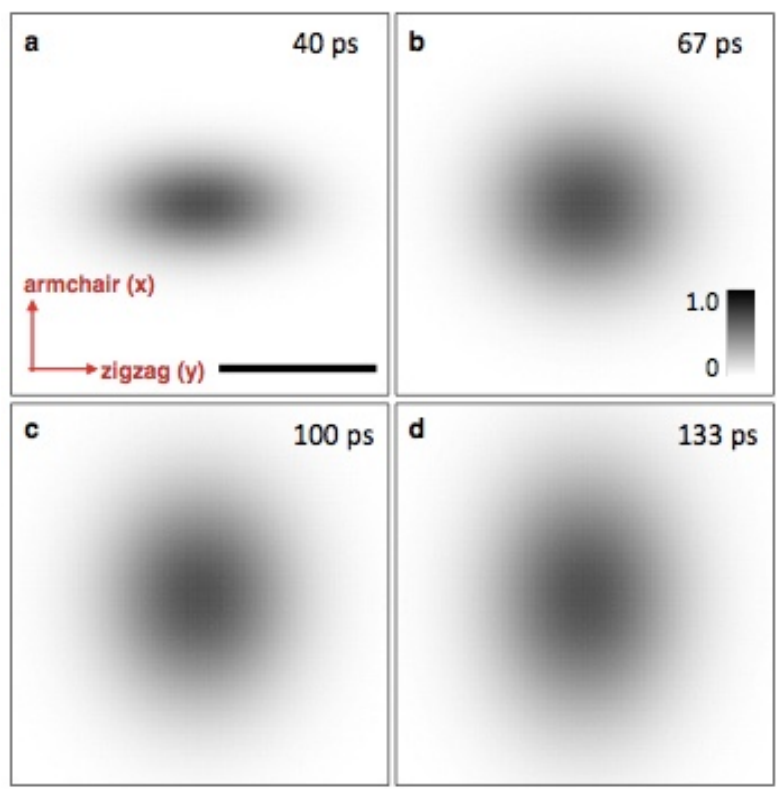

# Supplementary Information

# Spatial-Temporal Imaging of Anisotropic Photocarrier Dynamics in Black Phosphorus

Bolin Liao[1,2,⊥], Huan Zhao[3,⊥], Ebrahim Najafi[1], Xiaodong Yan[3], He Tian[3], Jesse Tice[4], Austin J. Minnich[2,5]*, Han Wang[3]* and Ahmed H. Zewail[1,2,§]

[1]Division of Chemistry and Chemical Engineering, California Institute of Technology, Pasadena, CA 91125, USA
[2]Kavli Nanoscience Institute, California Institute of Technology, Pasadena, CA 91125
[3]Ming Hsieh Department of Electrical Engineering, University of Southern California, Los Angeles, CA 90089, USA
[4]NG Next, Northrop Grumman, 1 Space Park, Redondo Beach, CA 90278, USA
[5]Division of Engineering and Applied Science, California Institute of Technology, Pasadena, CA 91125, USA

[⊥]These authors contributed equally.

*To whom correspondence should be addressed: han.wang.4@usc.edu (H.W.),

aminnich@caltech.edu   (A.J.M.)

[§] Deceased

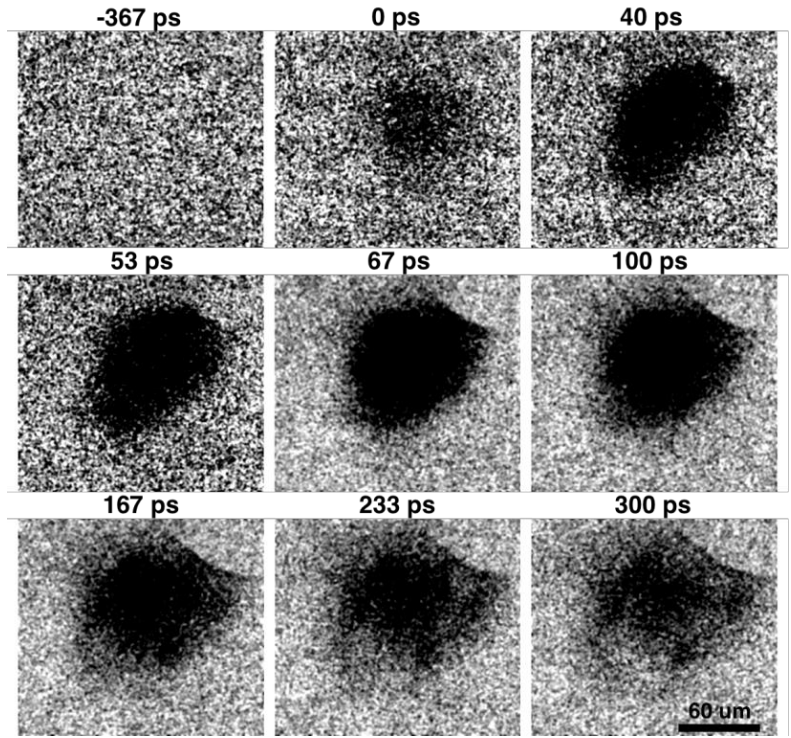

**Supplementary Figure 1. Hot hole dynamics in black phosphorus within 300 ps after photo-excitation.** This measurement was done on a different flake of black phosphorus from the one shown in the main text. The feature at the upper right corner is the boundary of the flake.

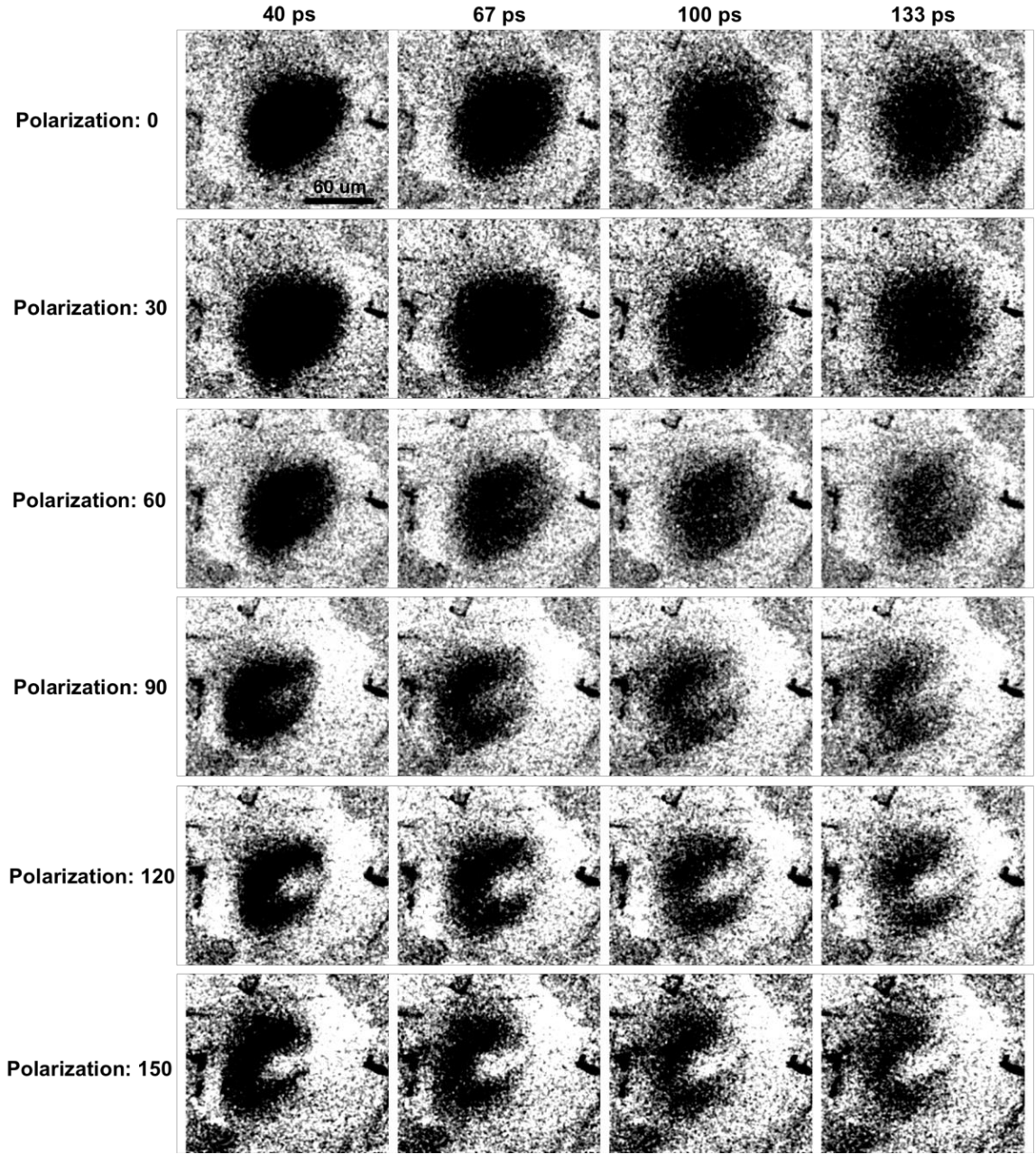

**Supplementary Figure 2. Polarization dependence of the hot hole dynamics in black phosphorus.** This measurement was done on a single BP flake with different polarizations of the input pump laser. Although the absorption of the pump laser is different with different polarizations (shown as different shapes and sizes of the induced hole distribution), the subsequent dynamics, namely the anisotropic diffusion, is the same for different polarization directions. The bright spot seen in the last three rows of images is a defect created due to long-time exposure of the sample to the pump laser.